\documentclass[acmtog]{acmart}

\AtBeginDocument{%
  \providecommand\BibTeX{{%
    \normalfont B\kern-0.5em{\scshape i\kern-0.25em b}\kern-0.8em\TeX}}}

\setcopyright{acmcopyright}
\copyrightyear{2023}
\acmYear{2023}
\acmDOI{XXXXXXX.XXXXXXX}

\acmJournal{TOSEM}
\acmVolume{37}
\acmNumber{4}
\acmArticle{111}
\acmMonth{8}

\begin{document}

\title{Using generative AI to support standardization work -- the case of 3GPP}

\author{Miroslaw Staron}
\email{miroslaw.staron@gu.se}
\orcid{0000-0002-9052-0864}
\affiliation{%
  \institution{Chalmers $|$ University of Gothenburg}
  \country{Gothenburg, Sweden}
}

\author{Jonathan Str{\"o}m}
\affiliation{%
  \institution{Ericsson AB}
  \city{Gothenburg}
  \country{Sweden}}
\email{jonathan.strom@ericsson.com}

\author{Albin Karlsson}
\affiliation{%
  \institution{University of Gothenburg}
  \city{Gothenburg}
  \country{Sweden}}

\author{Wilhelm Meding}
\affiliation{%
  \institution{Ericsson AB}
  \city{Gothenburg}
  \country{Sweden}}
\email{wilhelm.meding@ericsson.com}


\renewcommand{\shortauthors}{Staron et al.}

\begin{abstract}
Standardization processes build upon consensus between partners, which depends on their ability to identify points of disagreement and resolving them. Large standardization organizations, like the 3GPP or ISO, rely on leaders of work packages who can correctly, and efficiently, identify disagreements, discuss them and reach a consensus. This task, however, is effort-, labor-intensive and costly. In this paper, we address the problem of identifying similarities, dissimilarities and discussion points using large language models. In a design science research study, we work with one of the organizations which leads several workgroups in the 3GPP standard. Our goal is to understand how well the language models can support the standardization process in becoming more cost-efficient, faster and more reliable. Our results show that generic models for text summarization correlate well with domain expert's and delegate's assessments (Pearson correlation between 0.66 and 0.98), but that there is a need for domain-specific models to provide better discussion materials for the standardization groups. 
\end{abstract}

\ccsdesc[500]{Social and professional topics~Testing, certification and licensing}

\keywords{machine learning, standardization, requirements, automation}

\received{20 February 2007}
\received[revised]{12 March 2009}
\received[accepted]{5 June 2009}

\maketitle

\section{Introduction}
\noindent
Standardization in software and systems engineering is typically spearheaded by industry-led organizations such as the 3rd Generation Partnership Project (3GPP), the International Standards Organization (ISO) or Automotive AUTOSAR consortium. The development of new standards, or the modification of existing ones, relies heavily on reaching a consensus among member entities. Usually, the member organizations provide input before meetings, which needs to be read, organized, summarized, and then presented at the meetings \cite{baron2018unpacking, baron2019making}. 

In this context, sizable standardization companies face the time-consuming and effort-intensive task of analyzing a significant volume of documents that must be read, distilled, and showcased during discussions. Achieving consensus, while also delivering tangible benefits to the contributing bodies, necessitates an effective analysis of these submissions. A more streamlined process not only results in superior standards but also ensures a more effective use of the experts' time who are integral to the standardization journey. In the end, such a process has a significant impact on the society as a whole \cite{bruer2021mapping}. 

Currently, the analysis is mostly a manual task, with support only in terms of contribution change-tracking, but this can be improved with the use of Large Language Models (LLMs). We observe the rise of new LLMs that can help in the analysis of these large documents. Therefore, in this paper, we address the following research question:

\smallskip
\emph{To which degree can large language models provide relevant summaries of standardization documents?}
\smallskip

In particular, we set off to analyze the BART model and the Pegassus XLM models for the following tasks:
\begin{enumerate}
    \item summarization of large documents -- with the purpose of getting a quick orientation of the main contributions,  
    \item similarity analysis -- with the purpose of identifying the features which the members agree on (and where more discussions are needed), and
    \item overlap analysis -- with the purpose of guiding the discussions in smaller groups of companies. 
\end{enumerate}

We use the design science research approach \cite{wieringa2014design}, where the artifact is the machine learning-based system for supporting the standardization processes. We evaluate it with our industrial partner Ericsson AB in the context of their past 3GPP RAN standardization activities, following our previous studies \cite{ochodek2022chapter}. 

The rest of the paper is structured as follows. Section \ref{sec:related_work} outlines the most relevant related research studies in the area of text summarization and standardization efforts. Section \ref{sec:research_design} presents the details of our research design. Section \ref{sec:summarization} presents the artefact -- standards document summarization and analysis system -- and Section \ref{sec:results} presents the results of its evaluation. Section \ref{sec:discussion} discusses the results and the validity of our study. Finally, Section \ref{sec:conclusions} presents the conclusions. 

\section{Related work}
\label{sec:related_work}
\noindent
A lot of current research effort related to 3GPP standardization goes into the development of standards for UAVs (drones) \cite{abdalla2021communications}, where telecommunication plays a crucial role; in particular low latency communication. Standardization is important for all kinds of devices, not only drones \cite{kar2020critical}. The number of such standards indicates how much-automated support is needed to provide the technology development with standards that are up-to-date and relevant. The standardization is important for the market, but also guides the development of modern software during the process of decision making \cite{elliot2020artificial}.

Another line of research is related to the standardization process itself, for example, analyzing the cooperation vs. competition \cite{johansson2019research}. In particular the most interesting are studies of complex dependency networks between platforms, customers, suppliers, and infrastructure providers who can both compete in the same areas and collaborate in other areas \cite{ali2018complex, heikkila2023coopetition}. This coopetition (simultaneous cooperation and competition) has led to a surge in research studies that are \cite{gernsheimer2021coopetition}. However, this research is not focused on the support of processes for building consensus, but on the organizational topics, e.g., what drives tensions between companies or governance models. 

A few studies show the possibilities that large language models (like GPT -- Generative Pre-trained Transformers) provide for standardization. For example, SkillGPT \cite{li2023skillgpt} is effective in identifying a standard set of skills while searching for jobs. Despite the different domains, the approach is similar to ours -- instead of using complex feature engineering, the language models can summarize texts/skills embed them in latent space, and then compare them on a semantic level. Similarly, large language models were used to align software requirements w.r.t. writing style \cite{tikayat2023agile}. 

The language models were also used for extracting relevant information in a financial sector \cite{huang2023finbert}. Although done only for sentiment analysis, this study shows that large language models can be trained and used for domain-specific tasks and domain-specific data. In our work, we use the same architecture of models, although pre-trained on another type of data -- research papers and standardization documents. 

Recent advances in large language models technology show that these models perform better than humans in summarization tasks \cite{liu2023learning}, which is one of the elements of our pipeline. However, regardless of the model, most of the large language models perform similarly on summarization tasks \cite{zhang2023benchmarking}, when they are trained on similar data. However, GPT-3 and larger models perform much better when additional training is done, which is one of the further work directions in our study. 

\section{Research design}
\label{sec:research_design}
\noindent
Based on the fact that the language models perform well on similar tasks, but not in the domain of standard development or telecommunication, we designed a study to explore this further. Since we can work with a company that develops telecommunication infrastructure, as well as participate in the standardization efforts, we settled on design science research \cite{wieringa2014design}. It allows us to develop a prototype and evaluate it in vivo in the industrial context. Since our intervention is delimited to presentations and support of the industrial practitioners, we did not adopt action research \cite{staron2020action}. We designed the study to comprise of two cycles with the same company -- Ericsson AB -- and its unit working with 3GPP standardization. 

The starting point for the study was the industrial need to summarize and analyze contributions in the standardization context. In a real world scenario, the delegates summarize and analyze the contributions from all member organizations before and during the standardization meetings, which usually last for about a week.  

\subsection{Refined problem definition}
\noindent
In the first cycle, we addressed the problem of which machine learning model could be used for this task. We developed a prototype machine learning pipeline based on the XLM (Cross-Language Model) Pegasus to summarize member contributions. We evaluated it in a workshop with the delegate who worked with the standardization. 

\begin{figure*}[!htb]
    \includegraphics[width=\textwidth]{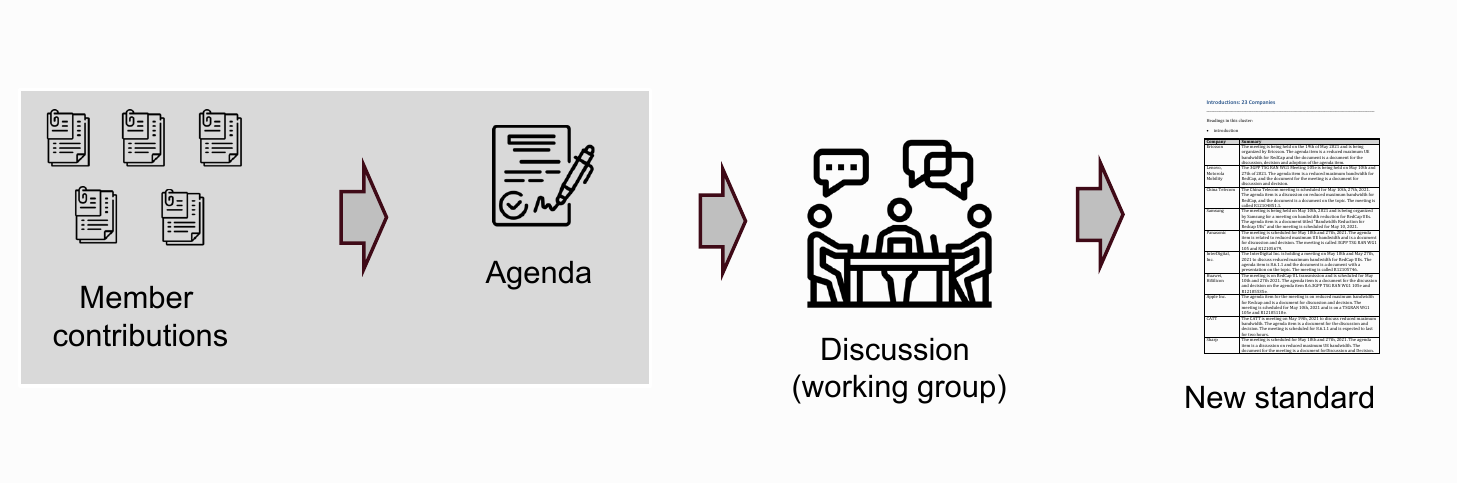}
    \caption{High-level overview of the working group meetings during the 3GPP standardization process. The grey background indicates the scope of this paper. }
    \label{fig:motivation}
\end{figure*}

The outcome was a refined problem formulation, where we identified the following workflow as the improvement (intervention) part of the process. This refined problem formulation can be conceptually presented in Figure \ref{fig:motivation}. The process usually starts with member companies (their representatives) submitting contributions to the meeting. These contributions can be lengthy documents and they need to be read, understood, and summarized by the leadership of the workgroup. The leader prepares the agenda for discussions in the working group. After the discussions, the working group reaches a consensus and this consensus becomes the standard (or part of the standard for which the working group is responsible).  

Although it is a straightforward process, the challenges are in the content of the contributions and in achieving the consensus. From the lengthy contributions, the leader needs to identify and extract information about:
\begin{itemize}
    \item items that all members agree on -- these will be presented in a summarized form during the meeting, 
    \item items that certain members do not agree on -- these need to be discussed in subgroups before, or during, the meeting, and
    \item items where the members' agreement is conditioned on certain changes, e.g., changing the allowed bandwidth -- these need to be discussed, agreed on, and modified during the meeting. 
\end{itemize}
\noindent
An example of an agenda item for discussion is presented in the box below (from 3GPP TSG-RAN WG1 Meeting \#105-e, in May 2021\footnote{\url{https://www.3gpp.org/ftp/tsg_ran/WG1_RL1/TSGR1_105-e/Docs/R1-2105999.zip}}): 
\newline
\newline
\noindent\fbox{%
    \parbox{\columnwidth}{%
        High Priority Proposal 3.1-1a:
        \newline
        \emph{Both during and after initial access, the scenario where the initial UL BWP for non-RedCap UEs is configured to be wider than the maximum RedCap UE bandwidth is allowed.}
    }%
}
\newline
\newline

\noindent
In the above text, the leader of the work package identified a change that should be done to the standard, based on the contributions from the members of the company. 

It is the effort- and labor-intensive process of preparing the summaries, identifying agreements, commonalities, and discrepancies, as well as summarizing the contributions that motivate our work. Since large language models (like GPT-4) have shown a significant potential for similar tasks for generic language tasks, we employ them in our design science research study. 

\section{Summarization and analysis system}
\label{sec:summarization}
In our Design Science Research (DSR) study, we developed the summarization and analysis system as the artifact. The system was designed to implement these three tasks:
\begin{enumerate}
    \item to summarize large documents to get a quick orientation of the main contributions,  
    \item to analyze the similarity of sections -- with the purpose of identifying the features which the members agree on (and where maybe more discussions are needed), and
    \item to analyze overlaps/similarities of entire documents -- with the purpose of guiding the discussions in smaller groups of companies. 
\end{enumerate}

The system supports the workflow of an individual contributor to the standard -- understanding of the content, identification of agreements and disagreements, and preparation of the agenda for further discussions. 

\begin{figure*}[!htb]
    \centering
    \includegraphics[width=\textwidth]{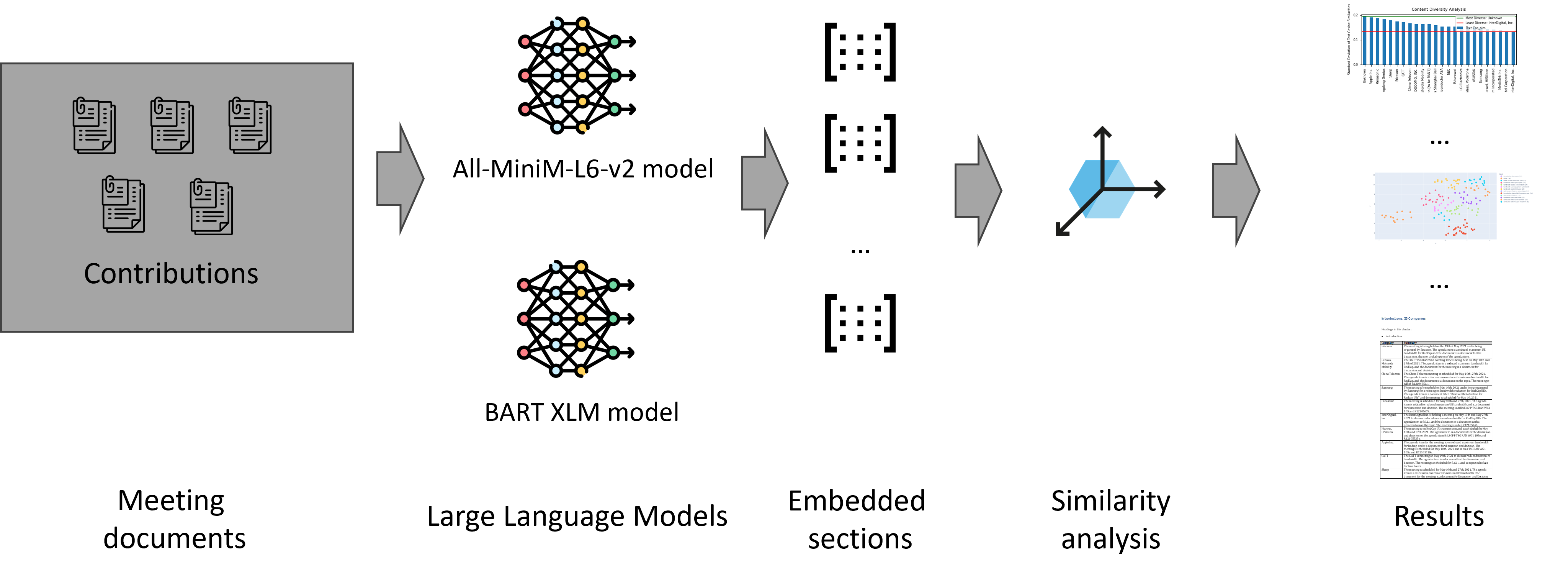}
    \caption{Artefact: System for analyzing and summarizing contributor documents}
    \label{fig:pipeline}
\end{figure*}

Figure \ref{fig:pipeline} presents this summarization and analysis system. 

The input to the system is the set of member contributions. These contributions are Microsoft (MS) Word documents which describe what each contributor company wants to discuss during the meeting -- agreements on the proposals, counter-proposals for alternative solutions, or plain disagreements when the company identifies a solution that cannot be adopted. 

The next step in our system is summarization of texts. We use the BART XLM model\footnote{\url{https://huggingface.co/docs/transformers/model_doc/bart}} which is open source cross-language model trained on technical texts and research articles. Our system extracts each heading from the contributor documents and uses the BART model to summarize them. These summaries are used later on to provide the user of the system to quickly understand the content of the documents. 

Then our system uses the All-MiniLM\footnote{\url{https://huggingface.co/sentence-transformers/all-MiniLM-L6-v2}} to extract embeddings from the text of each section (or subsection, depending on what the lowest level is) of each member document. We embed each sentence and then average the embeddings for the entire section. 

After extracting the embeddings, the system calculates the similarity/dissimilarity between these embeddings. We use the cosine distance to find the similarities, as it is an established similarity measure in natural language processing. To balance the content of each section and the heading, we also calculate the embeddings of the headings for each section (separate from the content of the section). When calculating the similarity we use a weighted average of the similarity between the content and the similarity of the heading. Our heuristic is based on the observation that the headings reflect the authors' intentions -- it captures the topic -- while the content reflects the details of this intention -- it captures the agreement/disagreement. 

\begin{figure}
    \centering
    \includegraphics[width=1\linewidth]{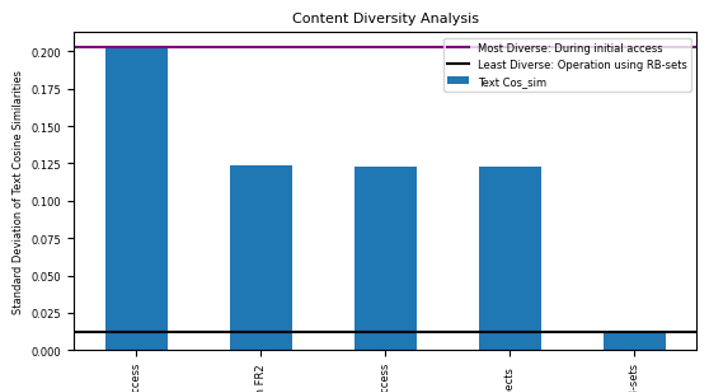}
    \caption{Part of a diagram showing similarity between sections in contributor documents. Each bar represents a pair of documents.}
    \label{fig:similarity_sections}
\end{figure}

\begin{figure}
    \centering
    \includegraphics[width=1\linewidth]{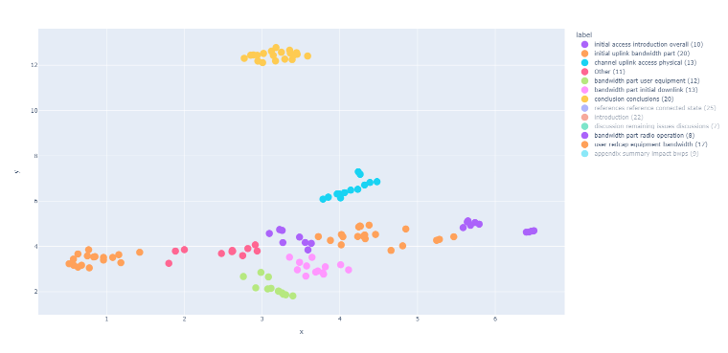}
    \caption{An example of clusters of content sections labeled with the most common words in the passages. Each dot represents a section on contributor documents. This diagram indicates which topics the contributors agree on the most.}
    \label{fig:similarity_tsne}
\end{figure}

The results from the analysis system are a set of diagrams and a proposal for discussion points for the agenda to the meeting -- presented in Figure \ref{fig:similarity_sections} -- Figure \ref{fig:similarity_tsne}.

First, the system provides an overview of the agreements on the section level. Since the number of pairs is quite large, it is useful to provide them as a long list of similarities and find the most and the least similar sections. We found that the barchart is the best diagram for that, with an example presented in Figure \ref{fig:similarity_sections}. 

For the visualization of the agreement between companies we also use barcharts and we complement them with a graph which shows how the contributors cluster with each other. 


Finally, it is important to understand the topics on which the contributors agree. So the next diagram is the visualization of clusters of topics, presented in Figure \ref{fig:similarity_tsne}. We use the t-SNE dimensionality reduction technique to reduce the embedding vectors of 768 elements to two dimensions.  

The agenda includes a summary of the topics discussed and the identification of potential agreements and disagreements.  

\subsection{Evaluation}
\noindent
We evaluated our approach through a workshop with the company's delegate who was a part of the 3GPP standardization process and was leading the meeting that we analyzed. The goal of the evaluation was to test how useful the support of the model-generated reports is. For the evaluation, we set off to address the following research questions in particular: 

\begin{enumerate}
    \item To which degree is the summarized text reflecting the technical content of the document?
    \item To what degree can the model identify agreements in member contributions?
    \item To what degree can the model identify agreements between member companies?
\end{enumerate}

\noindent In the first question, we assessed to which degree the technical content is captured by the model. The technical content of the standardization can be quite detailed -- e.g., one member may request the bandwidth to be 20 Mhz while another one to be 40 Mhz. This technical detail is crucial but may be hidden in a larger portion of the text in the members' contributions. 

In the second question, we assessed the degree of agreement between the contributions. A consensus is built in such meetings by analyzing the differences, understanding one another, and agreeing on the way forward. This means that the wording of the contributions is important here. For example, one member's contribution can talk about "accepting 20 Mhz" while another one is about "not accepting 20 Mhz" -- again, a tiny change in wording can make a big difference. 

In the third question, we assessed the degree of consensus between companies, i.e., members of the consortium. This is an important part of the standardization as the member companies represent different sectors that have stakes in the standardization. Infrastructure providers have a different focus than equipment manufacturers or chipmakers. Identifying agreements across companies, and within sectors, helps to drive the standardization forward in terms of finding the parts of the standard that mature faster than other parts of the standard. For example, infrastructure providers may agree upon the base technology faster while the consumer product manufacturers need to discuss the details of the rollout of the technology to the customers. 

Finally, all of these questions assess the degree, not only existence/non-existence of agreement, because in the standardization meeting, the members can voice their opinions, and therefore the output of the model should provide good support for the discussion, not automatically replace these discussions. Using the degree of agreement allowed us to have a nuanced discussion with the delegate. 

\subsubsection{Analysis 1: Quality of summaries}
\noindent 
To make the comparison relative and fair we use the following method: 
\begin{itemize}
    \item We randomly select 10 summaries.  
    \item We ask the expert (and the delegate) to rate them on a scale of 1-5 -- since the similarity rank provided by the ML model is in the form of an index, we also choose the expert (and the delegate) to grade the similarity on the ordinal scale.  
    \item We ask the expert (and the delegate) to explain his assessment -- to understand the reasoning behind the assessment and to understand what kind of information was captured correctly/incorrectly by the model. 
\end{itemize}
\noindent
The analysis is qualitative as we are interested in understanding the limitations of the algorithms. We capture the assessment in three categories. Form and structure is where we ask the expert and the delegate to assess whether the text of the summary reflects the structure of the original text -- for example whether vital information that is included in bullet points is somehow reflected in the summary. Content is the category where we ask the expert and the delegate to assess whether the summary captures the content sufficiently, e.g., whether vital information is included. The domain category is where the expert and the delegate both assess whether the summary captures information that is important for the domain, e.g., change of a bandwidth from 100MHz to 20MHz. 

\subsubsection{Analysis 2: Quality of agreement assessment per agenda item}
When we assess the quality of the agreement between the expert or the delegate and the algorithm, we use the following method based on calculating the correlation coefficient, similar to Antinyan et al. \cite{antinyan2017rendex}.

\begin{itemize}
    \item We select the top 5 agreements, where the algorithm calculated the similarity between agenda items (sections in the contribution document) to be the highest.  
    \item In addition to that we select the bottom 5 agreements, where the agreement is the lowest, to balance the assessment. 
    \item We ask the expert (and the delegate) to rate them on a scale of 0 -- 1. 
    \item We ask the expert (and the delegate) to explain his assessment and to understand the reasoning behind the assessment. 
    \item We calculate the Pearson correlation coefficient to quantify the strength of the agreement between the expert's and the delegate's assessments and the algorithm. 
\end{itemize}

\noindent
This analysis combines the quantitative assessment and the qualitative one. In addition to the number that quantifies the agreement's strength, we also need to understand how the expert/delegate reasons when comparing the texts. This helps us to improve the algorithm in the future. 

\subsubsection{Analysis 3: Quality of agreement assessment per contribution} In the final stage of the analysis we raise the level of abstraction and analyze the entire contribution documents -- compared to Analysis 2. 

We follow a similar process: 
\begin{itemize}
    \item We select the top 5 agreements of the entire contributions. 
    \item We select the bottom 5 agreements of the entire contributions. 
    \item We ask the expert/delegate to rate them on a scale of 0 -- 1. 
    \item We ask the expert/delegate to explain his assessment. 
    \item We calculate the Pearson correlation coefficient. 
\end{itemize}

\noindent
This analysis focuses on comparing the entire documents, as we need to understand how feasible it is to quickly get an orientation about similarities between contributions (and therefore the companies). 

\section{Results}
\label{sec:results}
For the evaluation of the approach, we chose publicly available data from the 3GPP standardization committee. We selected one meeting, for which we analyzed both the summary of the meeting and the contributions from the member companies\footnote{Meeting R1-2105999, TSG-RAN WG1 Meeting \#105-e. The document is available at: \url{https://www.3gpp.org/ftp/tsg_ran/WG1_RL1/TSGR1_105-e/Docs}}. We selected the meeting since our expert evaluator was part of that meeting and could provide detailed insights, which in turn allowed us to understand the quality of our approach. 

We collected data from one domain expert and one company's delegate in the 3GPP consortium. The first one is a domain expert who does not participate in the standardization meeting. The first analysis is therefore called the pilot analysis as we mostly focus on evaluating the methodology. Since the number of delegates who lead the standardization meeting is limited, we decided to ask one of them for the final evaluation.

\subsection{Visualization of similarities}
\begin{figure*}[!tb]
    \centering
    \includegraphics[width=0.8\textwidth]{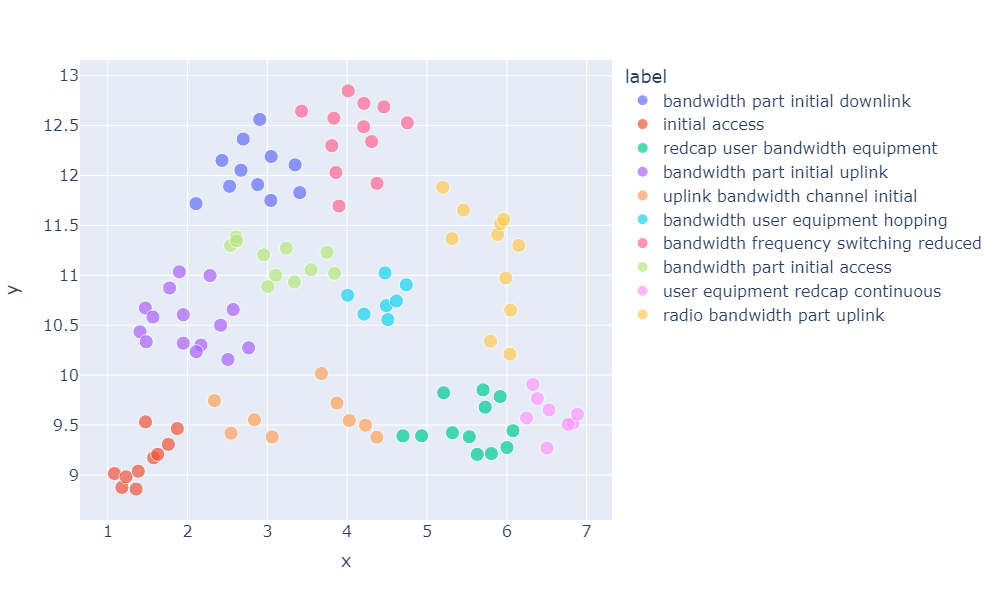}
    \caption{Visualization of the headings and content of each paragraph of the contributions. Each dot represents one section from one document. Each color represents one cluster (based on the k-Means clustering algorithm with k=10).}
    \label{fig:clusters}
\end{figure*}

\noindent
We used the cosine similarity to find and cluster headings and the content of the contributions. Figure \ref{fig:clusters} presents a t-SNE transformed diagram where each cluster represents a specific topic -- the label. 

Each cluster is labeled using the top common concepts that are used in the text. 

The concept/word "bandwidth" is used repeatedly and rightfully so, since these texts were submitted to a meeting that discussed the bandwidth parts standardization -- for both uplink and downlink. 

Grouping of the topics provides a basic understanding of which groups of topics are discussed. Using the t-SNE diagram allows us to visualize the topics, but we cannot assess how close or far away the topics are from one another, or whether the contributors agree or disagree about these topics. Therefore, we can use the cosine distance and visualize the distribution of the distances between all pairs of headings, as well as all pairs of contents under these headings. 

\begin{figure}
    \centering
    \includegraphics[width=\columnwidth]{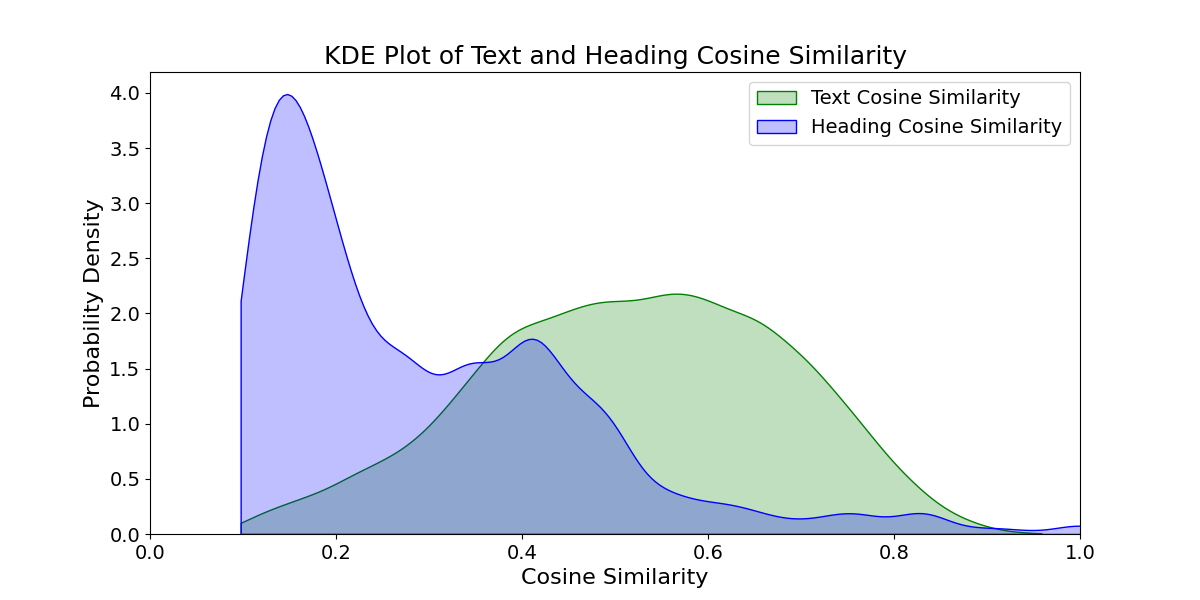}
    \caption{Distribution of cosine similarities between contents of the contribution. Each heading was transformed into a feature vector and compared to all other headings; the same was done for the content of each section. }
    \label{fig:cosine_distribution_kde}
\end{figure}

The distribution of cosine similarities (pairwise) between all headings (blue color) and all content (green color) is presented in Figure \ref{fig:cosine_distribution_kde}. The diagram shows that the distribution of headings has one distinct peak around 0.1, which means that the headings are often describing different aspects. This is according to expectations as the headings are both shorter and also must succinctly characterize the content. The content, however, is distributed more according to the normal distribution. This is expected as the content is often a more elaborate text, including more similar words (e.g., technical terms like "MHz" and "bandwidth"), which are used in different sentences. 

\begin{figure}
    \centering
    \includegraphics[width=\columnwidth]{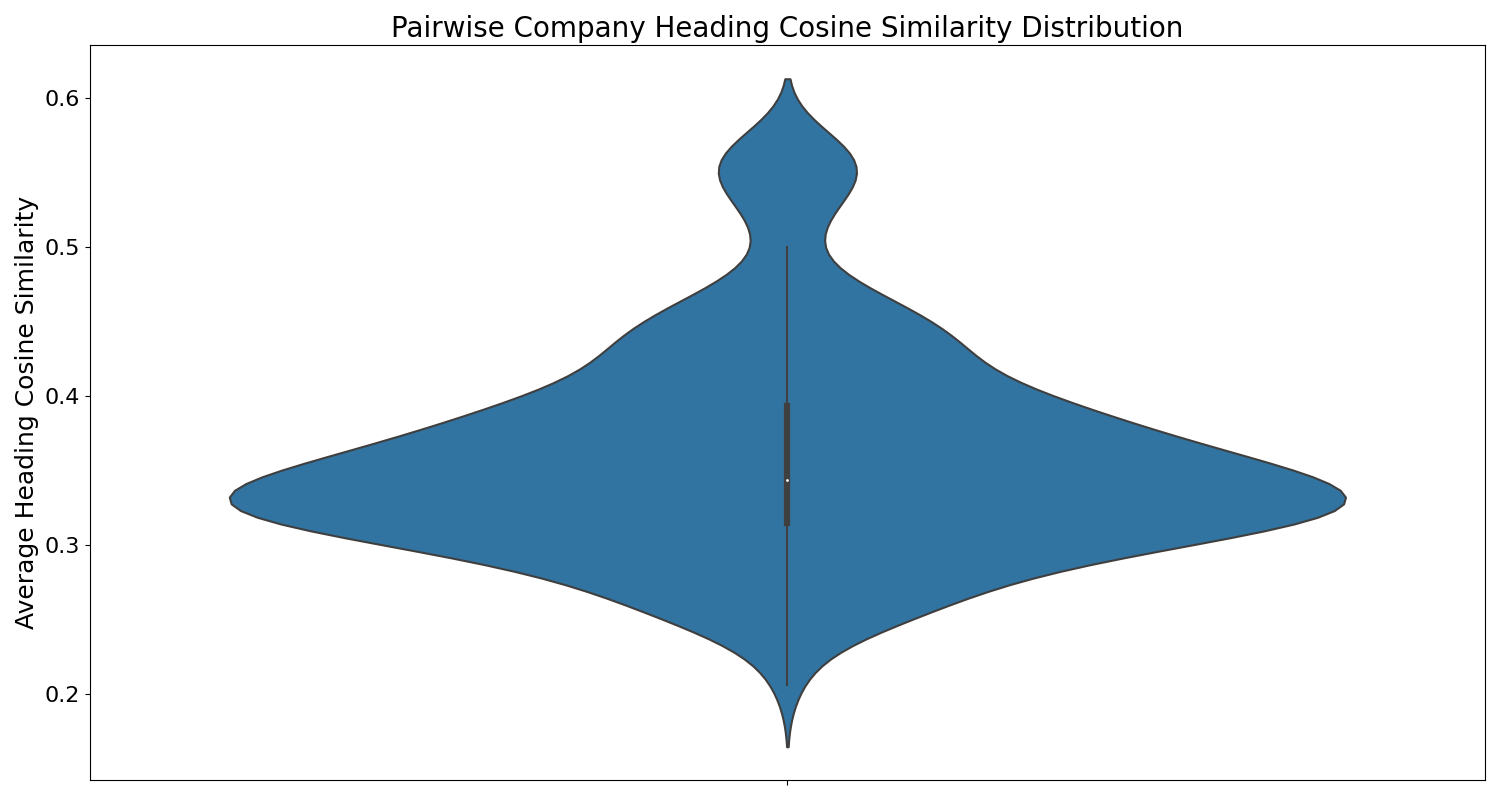}
    \caption{Distribution of the cosine similarity between pairs of headings in all sections in the contribution documents. }
    \label{fig:violin_heading}
\end{figure}

\begin{figure}
    \centering
    \includegraphics[width=\columnwidth]{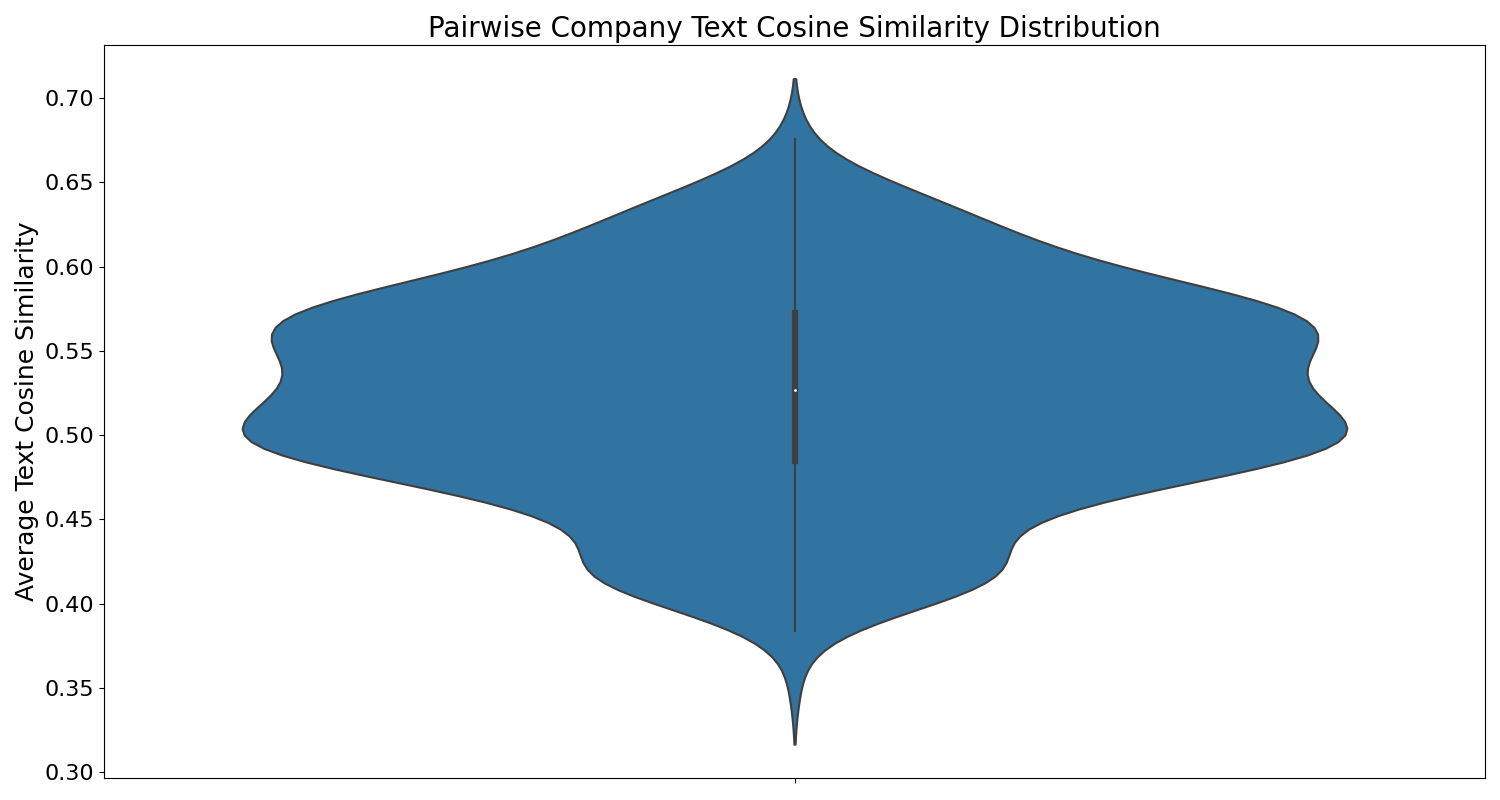}
    \caption{Distribution of the cosine similarity between pairs of the content of all sections in the contribution documents. }
    \label{fig:violin_content}
\end{figure}

These distributions are even more visible when we plot them on violin diagrams in Figure \ref{fig:violin_heading} and Figure \ref{fig:violin_content}. We should note here that the distributions are concentrated around 0.1 and 0.5 for the headings and content respectively. There are no data points that are on the negative side of cosine similarity (i.e., below 0.0), indicating that the topics are opposite to one another. 

Since these distributions are so different, when selecting the similar and dissimilar passages, we combined the similarity of the content and the heading (with 50\% weights).

\subsection{Pilot analysis -- quality of summaries}
The first analysis is the assessment of the quality of the summaries. The analysis is presented in Table \ref{tab:summary_quality_pilot}. The summaries are not included in the paper but can be accessed in our replication package. 

\begin{table*}[!htb]
    \centering
    \footnotesize
    \begin{tabular}{||p{4.8cm}|c|c|c|c|c|p{9cm}||}
        \hline Title&Sum.&Orig.&F/S&C&D&Comment \\ \hline
                On reduced max UE bandwidth for RedCap&1.1&1.2&1&2&3&Original text 5 pages. The algorithm failed to give a proper summary and to divide the text into DL and UL sub-parts\\ \hline
                Reduced maximum UE bandwidth for RedCap&2.1&2.2&1&2&3&Original text 5 pages. The algorithm failed to give a proper summary and to divide the text into DL and UL sub-parts. \\ \hline
                Discussion on reduced maximum UE bandwidth for RedCap  &3.1&3.2&4&2&3&Erroneous summary, e.g., it mentioned DL, which is not mentioned in the text. It did not include the main sum-up of the original content. The text is within the broader domain. but does not address the details. \\ \hline
                \textbf{Aspects related to reduced maximum UE bandwidth}&4.1&4.2&4&4&5&The algorithm captured the essence of the content, but not nuances like boldface text. \\ \hline
                Reduced maximum bandwidth for RedCap UEs&5.1&5.2&3&3&5&The algorithm captured the essence of the content but failed to provide the three options listed in the summary, which are the main contribution of the content. \\ \hline
                \textbf{Discussion on reduced maximum UE bandwidth for RedCap}&6.1&6.2&4&4&5&The algorithm captures the essence of the content, but with a poor syntax, making it difficult to grasp what it is saying. \\ \hline
                On reduced maximum UE bandwidth for Redcap&7.1&7.2&3&3&5&The algorithm captures one of the two parts of the content; the summary and the language are ok. \\ \hline
                UE Complexity Reduction Aspects Related to Reduced Maximum UE Bandwidth&8.1&8.2&3&2&4&The algorithm captures parts of the content. It has difficulties in handling proposals. \\ \hline
                Ensuring coexistence between RedCap and non-RedCap UEs  &9.1&9.2&5&2&4&The algorithm misses the essence of the content. which in this case is that by lowering the frequency from 100MHz to 20MHz power savings can be made. What it writes is though correct. but not complete and not the main part of the content. \\ \hline
                Discussion on Bandwidth Reduction for RedCap UEs&10.1&10.2&3&3&5&The algorithm should have mentioned uplink also; it mentions vital parts of the content but does not conclude the sentence. \\ \hline
    \end{tabular}
    \caption{Assessment of the quality of the summaries. The headers are sum -- a reference to the paragraph with the summary, orig. reference to the original text, F/S -- quality of the form and structure of the text on a scale of 1--5, C -- quality of the content, D -- quality of the domain-specific aspects in the text, Comment -- comment from the expert.}
    \label{tab:summary_quality_pilot}
\end{table*}

In the table, we highlight two summaries that are ranked as the most similar across all three categories -- 3, 4, and 5. The expert commented that the algorithm missed some of the nuances in the text, which is rather acceptable as the summary is not the same as the full document. 

On the other hand, the first two rows are ranked very low by the expert. These two summaries are ranked low because of the combination of the length of the text and many technical details entangled in the text -- discussions of downlink (DL) and uplink (UL). 

A short analysis of the qualitative feedback from the expert shows that the summaries are dependent on the length of the text. The longer the text, the worse the summary, something that is supported by the existing body of knowledge \cite{mani2002summac}. 

\subsection{Pilot analysis -- quality of agreement assessment per agenda item}

Firstly, Table \ref{tab:similarity_heading_pilot} presents the results of the similarity and the expert's assessment with his comments. The table includes both the five best and five worst pairs as ranked by the similarity assessment algorithm. 

\begin{table}[!htb]
    \centering
    \footnotesize
    \begin{tabular}{||c|l|c|p{5cm}||} \hline
        Pair & Algorithm & Expert & Comment \\ \hline
        PA-1 & 0.92 & 0.9 & Mismatch in text -- Uplink vs. uplink + downlink. \\
        PA-2 & 0.91 & 0.8 & Content are similar, but one of the reports is more detailed. \\ 
        PA-3 & 0.91 & 0.7 & Content are similar, but one of the companies also takes up additional topics. \\
        PA-4 & 0.91 & 0.7 & Same comment as above. \\
        PA-5 & 0.9 & 0.8 & Same comment as above. \\ \hline
        NA-1 & 0.11 & 0.2 & The same topic, but from different perspectives. \\
        NA-2 & 0.12 & 0.1 & Only one common aspect, otherwise the topics are different. \\
        NA-3 & 0.12 & 0.2 & Only one common aspect, otherwise the topics are different. \\
        NA-4 & 0.12 & 0.1 & Same as above. \\
        NA-5 & 0.13 & 0.2 & Same as above. \\ \hline
    \end{tabular}
    \caption{Assessment of the similarity per pair. PA -- positive match, NA -- negative match, Algorithm -- model's similarity scope (between 0 and 1); Expert -- expert's similarity score (between 0 and 1); Comment -- summary of the expert's explanation about the similarity and difference. }
    \label{tab:similarity_heading_pilot}
\end{table}

When we calculate the Pearson correlation coefficient, we see a high correlation of 0.98. This is an indicator that even if the summary (in the previous analysis) is not perfect, the algorithm still manages to provide a good match when finding similar sections in documents. 

The qualitative analysis indicates that there is one common mistake that the algorithm makes -- it misses the fact when one of the sections is longer and therefore includes additional information. 

At the same time, when two summaries discuss topics from different perspectives, the algorithm ranks such summaries as different from one another. This indicates that the results of the similarity calculation are correct. 

\subsection{Pilot study -- quality of agreement per agenda item}
Table \ref{tab:similarity_contribution_pilot} presents the expert's and the algorithm's similarity assessment for the entire documents. The method is similar to the previous analysis, except that here the input is the entire contribution documents. The size of these documents can vary from a few pages (5) to many pages (30). 

\begin{table}
    \centering
    \footnotesize
    \begin{tabular}{||c|l|c|p{5cm}||} \hline
        Pair & Algorithm & Expert & Comment \\ \hline
        PC-1 & 0.6 & 0.6 & Both companies talk about the same thing, but one is more detailed. \\
        PC-2 & 0.58 & 0.8 & Same as above. \\ 
        PC-3 & 0.57 & 0.8 & Same as above. \\
        PC-4 & 0.57 & 0.7 & Same as above. \\
        PC-5 & 0.57 & 0.6 & Same as above\\ \hline
        NC-1 & 0.34 & 0.4 & One of the companies had commonalities with several other companies. \\
        NC-2 & 0.32 & 0.8 & One of the companies was much more detailed. \\
        NC-3 & 0.32 & 0.3 & One report has a very narrow focus (uplink only). \\
        NC-4 & 0.3 & 0.3 & Same as above. \\
        NC-5 & 0.28 & 0.7 & Many diagrams in one of the reports. \\ \hline
    \end{tabular}
    \caption{Assessment of the similarity per pair of reports/contributions. PC -- positive match, NC -- negative match, Algorithm -- model's similarity scope (between 0 and 1); Expert -- expert's similarity score (between 0 and 1); Comment -- summary of the expert's explanation about the similarity and difference. }
    \label{tab:similarity_contribution_pilot}
\end{table}

The Pearson correlation coefficient is 0.49 which indicates lower agreement between the expert and the algorithm. In closer scrutiny, we can see that the algorithm's similarity assessment is not as widely spread as the experts -- it ranges from 0.28 to 0.6, while the expert's assessment ranges from 0.3 to 0.8. This is caused by the size of the documents -- larger documents' embeddings draw more towards the average and therefore their similarity is higher. 

When exploring the quantitative comments, it shows that the level of detail in the documents is important -- resulting in higher similarity. The presence of figures (images) in the documents is also important and not captured by the algorithm. Since the algorithm designed in this study is for text documents, this is natural and will result in further improvement of our toolkit. 

\subsection{Changes based on the pilot study}
\noindent
In the pilot study, we identified the fact that the algorithm misses important information in the form of proposals and scenarios. These are often dedicated parts of the text which were treated in the same as other parts of the document. Therefore, before showing the summaries to the second expert for the final evaluation, we added a simple counter which listed the scenarios and proposals present in the analyzed text. This allows the expert to get an orientation about the trustworthiness of the summary and allows the expert to focus on the text where scenarios and proposals are visible. 

\subsection{Evaluation with the delegate}
\noindent
For the first task, the delegate provided his summaries in a different format than in the pilot study. He provided detailed comments on the summaries in the document. His general comment was that the quality of the summaries varied a lot, but the majority of the summaries captured basic elements and not the right level of detail. 

His evaluation strengthened the view from the pilot study that the summaries should emphasize (and list) proposals presented in the documents; and should put less focus on the text. He also identified misplaced summaries -- technical questions were summarized in a place that was dedicated to more general issues. For example, text about uplink was summarized in the context of a text discussing downlinks. When scrutinizing the text we found that there was indeed uplink mentioned in the text, but it was not the main topic -- the model attached the attention to the wrong topic. 

For the second task, the delegate provided the full set of answers, presented in Table \ref{tab:similarity_heading_expert}. 

\begin{table}[!htb]
    \centering
    \footnotesize
    \begin{tabular}{||c|l|c|p{5cm}||} \hline
        Pair & Algorithm & Delegate & Comment \\ \hline
        PA-1 & 0.92 & 0.45 & One of the companies takes up topics that are not covered by the other company. \\
        PA-2 & 0.91 & 0.35 & Hard to compare both contributions; both take up multiple aspects, but they also differ a lot. \\ 
        PA-3 & 0.91 & 0.45 & Similar comment to the first pair PA-1. \\
        PA-4 & 0.91 & 0.4 &  \\
        PA-5 & 0.9 &  0.35 & Similar comments to PA-2 \\ \hline
        NA-1 & 0.11 & 0 & Different topics are covered \\
        NA-2 & 0.12 & 0 & Same as above \\
        NA-3 & 0.12 & 0 & Same as above \\
        NA-4 & 0.12 & 0 & Same as above \\
        NA-5 & 0.13 & 0 & Same as above \\ \hline
    \end{tabular}
    \caption{Assessment of the similarity per pair. PA -- positive match, NA -- negative match, Algorithm -- model's similarity scope (between 0 and 1); Delegate -- delegate's similarity score (between 0 and 1); Comment -- summary of the delegate's explanation about the similarity and difference. }
    \label{tab:similarity_heading_expert}
\end{table}

The correlation between the expert's assessment and the model is 0.98, but it needs to be treated with caution. The high correlation is partially caused by the 0s in the table. A more important aspect is the fact that the delegate clearly states that the negative matches are dissimilar -- 0 in the assessment. 

According to our protocol, we also asked the delegate to provide us with an assessment of the agreements and disagreements between the entire contributions. The results are presented in Table \ref{tab:similarity_contribution_expert}. 

\begin{table}
    \centering
    \footnotesize
    \begin{tabular}{||c|l|c|p{5cm}||} \hline
        Pair & Algorithm & Delegate & Comment \\ \hline
        PC-1 & 0.6 & 0.15 & Company 1 covers many more topics than Company 2.  \\
        PC-2 & 0.58 & 0.3 &  \\ 
        PC-3 & 0.57 & 0.3 &  \\
        PC-4 & 0.57 & 0.2 &  \\
        PC-5 & 0.57 & 0.2 & \\ \hline
        NC-1 & 0.34 & 0 &  \\
        NC-2 & 0.32 & N/A &  \\
        NC-3 & 0.32 & N/A &  \\
        NC-4 & 0.3 &  0.15 &  \\
        NC-5 & 0.28 & N/A &  \\ \hline
    \end{tabular}
    \caption{Assessment of the similarity per pair of reports/contributions. PC -- positive match, NC -- negative match, Algorithm -- model's similarity scope (between 0 and 1); Delegate -- delegate's similarity score (between 0 and 1); Comment -- summary of the delegate's explanation about the similarity and difference. }
    \label{tab:similarity_contribution_expert}
\end{table}

There, the correlation is 0.66, but it also needs to be taken into consideration cautiously, as the delegate did not provide the scores for three of the pairs. His assessment of the similar pairs, however, was much lower than the assessment of the pilot study expert. As this delegate was part of the meeting and knew the details of the agenda and the discussions, he could see even small details of the text that the model missed in the similarity. 

This finding is important as it indicates that this approach, and the model, can provide support for delegates, but cannot replace them. The details of the text are still very important. 

The overall comment is that it's difficult to compare similarities on the document level, because of the diversity of topics covered. It is better to compare the documents topic-by-topic. The delegate was also clear in his feedback that this is a very promising approach and could save a significant amount of effort in the context of standardization, where consensus-building is very important. 

\subsection{Summary of the evaluation}
\noindent
To summarize the evaluation, we found that the approach is effective in helping the standardization. The summaries provide a good way to get the initial orientation in the topics and in the agreements/disagreements between the contributions. 

The challenges with the approach are in the detailed analysis of documents. The publicly available models are not specific for the 3GPP standardization documents and therefore sometimes miss the intricacies of the text from this domain. 

Both the expert and the delegate in our evaluation acknowledge the value of this approach and indicate the need to use dedicated models that can provide better domain-specific summaries. 

\section{Validity analysis}
\label{sec:discussion}
Our study has been done in an industrial context, which has its specific validity threats, which we considered based on the frameworks of Wohlin \cite{wohlin2012experimentation} which is a de-facto standard in software engineering and Staron \cite{staron2020action}, which targets industrial contexts. Our choice of the research method and the approach prioritized choices that led to increased external validity. 

Regardless of our choice to optimize towards \emph{external validity}, we still see a limitation of the generalizability of our results. We focused on the text-based standardization process, as opposed to diagram- or model-based like OMG (Object Management Group). Analyzing diagrams is different and we plan to address that in our next study. 

Our main threat to the \emph{construct validity} is related to the selection of the document. Although the 3GPP standardization has a significant number of documents, we chose one of them where the delegate was involved. It allowed us to perform an in-depth evaluation of the content. Despite a bit longer time between the meeting (in 2021) and the study (2024), the delegate was able to provide insights without experiencing the maturity or the historical effects. 

The main threat to the \emph{conclusion validity} is the fact that we performed quantitative analysis, with a limited number of persons. One of the authors was part of the ISO standardization committee before the study and we had one domain expert and one delegate in the study to reduce the risk of bias in this study. However, we plan more studies in the next steps. 

Finally, we also identified an \emph{internal validity} threat in the form of a mono-operation bias -- we used only one approach and compared it to manual summary in the expert/delegate assessment. We relied on the expert's and the delegate's assessments for similarity analysis, not on automated distance measures, because the summaries done manually do not necessarily form an oracle (there are multiple ways of expressing the information).

\section{Conclusions}
\label{sec:conclusions}
Though time-intensive, standardization efforts are crucial for the development of sustainable, enduring, and secure software systems. However, the current standardization framework often faces limitations due to the constrained capacity of member organizations to manage the extensive information exchanged during meetings. 

The findings of this study highlight the utility of large language models in providing relevant summaries, yet indicate areas for improvement. While these summaries aid in initial comprehension and orientation within standardization groups, they fall short of effectively extracting detailed technical information, requiring manual intervention. Consequently, while these models offer valuable support without additional pre-training, their full potential can be achieved with additional, domain-specific, pre-training, which is the subject of our current work. 

The summaries of the content are better on topic, heading, and levels rather than on the document level. Although this generates more data that needs to be processed, it can lead to more detailed discussions. 

We also identify the following directions of our current research study: 
\begin{itemize}
    \item Train the model of 3GPP documents to capture the domain better -- the goal is to increase the sensitivity of the model to pinpoint the points of disagreements/discussions. 
    \item Include image analysis pipeline, to provide extra information about the certainty of the summaries -- e.g., many images should indicate that the summary is not certain as we miss a lot of information. 
\end{itemize}

\begin{acks}
This research has been partially financed by Software Center, Ericsson AB, University of Gothenburg -- \url{www.software-center.se}.
\end{acks}

\bibliographystyle{ACM-Reference-Format}
\bibliography{sample-base}

\end{document}